\documentclass{article}
\usepackage[utf8]{inputenc}
\usepackage{graphicx}
\usepackage{multirow}
\usepackage{amsmath}

\title{Secure Computation in Decentralized Data Markets}
\author{Fattaneh Bayatbabolghani and Bharath Ramsundar\\
Computable}

\begin{document}

\maketitle

\begin{abstract}
Decentralized data markets gather data from many contributors to create a joint data cooperative governed by market stakeholders. The ability to perform
secure computation on decentralized data markets would allow for useful insights to be gained while respecting the privacy of data contributors. In this paper, we design secure protocols for such computation by utilizing secure multi-party computation techniques including garbled circuit evaluation and homomorphic encryption. Our proposed solutions are efficient and capable of performing arbitrary computation, but we report performance on two specific applications in the healthcare domain to emphasize the applicability of our methods to sensitive datasets.  
\end{abstract}

\section{Introduction}
One of the challenges of building a decentralized data market \cite{ComputableWhitepaper} is providing adequate protection for the privacy of data contributors. Data contributors might be unwilling to contribute sensitive information into a data market if they lack adequate protections for their data. Economic considerations may ease some of these worries, but for high-value datasets more powerful cryptographic tools may be necessary to secure user data.

In this paper, we introduce a scenario where different data contributors (makers) wish to share their data (listings) to make data available for buyers who wish to perform specific computations on aggregated data. We assume makers are not comfortable sharing plaintext data. Therefore, our main goal in this work is performing computation on protected and aggregated data. In many examples in practice, these listings are not necessarily physically stored in one database (datatrust) or are not always owned by one organization. 

In previous work we introduced decentralized data markets \cite{computable, ComputableWhitepaper} which provide a powerful framework for constructing datasets with distributed ownership and control. We also introduced the maker/listing/datatrust terminology which we will reuse in this current paper. In these scenarios, storing protected listings and performing computation on them is not straightforward. Who performs encryption upon listings? How is computation done on encrypted data? In this paper, we explore these questions on two healthcare inspired examples: performing logistic regression on the breast cancer Wisconsin Dataset \cite{dataset}, and the linkage disequilibrium test on GWAS data. We design a secure solution to compute both logistic regression and linkage disequilibrium tests, but our designed protocol is general and can be used to perform arbitrary computation.  

In the following sections, we first provide some background related to the computation of logistic regression \cite{kleinbaum2002logistic}, linkage disequilibrium \cite{pritchard2001linkage}, and other cryptographic tools \cite{sealcrypto,yao1982protocols,yao1986generate}. Then we introduce our designed protocols, and at the end provide our experimental results. 

\section{Background}
In this paper we study two sample computational problems: logistic regression (LR) and the on linkage disequilibrium (LD) test performed genome-wide association study (GWAS) data. We design protocols for performing these computations on encrypted data and base on two cryptographic techniques: Homomorphic Encryption (HE) and Garbled Circuit (GC). In the following, we briefly provide needed background before moving to the design of our proposed protocol.

\subsection{Logistic Regression}
Logistic regression is a statistical method for analyzing a dataset in which there are one or more independent variables that determine an outcome. In this paper, our focus is on binary LR where LR is used to predict the relationship between independent variables and a dependent variable where the dependent variable is binary. We can divide the computation of LR into two categories: training and testing. In training, a model is trained based on training samples and parameters are computed. In testing, the trained model is applied on test cases. There are some standard open source tools to perform the training phase of LR such as TensorFlow \cite{abadi2016tensorflow}, PyTorch \cite{paszke2017automatic}, and SKLearn \cite{pedregosa2011scikit}. For the rest of this paper, we assume that we have access to a trained LR model and all its learned parameters and only focus on the implementation of LR for the testing phase. 

For LR during the testing phase, if we have fixed dimension $n$, precomputed parameters $W = (w_1, \ldots, w_n)$ and $b$ ($W$ and $b$ are regression coefficients of a trained model), and a sample $X = (x_1, \ldots, x_n)$, then we can compute probability $p$ as:
$$p = \frac{e^{X\cdot W+b}}{1-e^{X\cdot W+b}}$$
As you can see in the computation of $p$, if we can compute $e^{X\cdot W+b}$, we can easily compute the rest of the probability. We focus on computing this quantity on encrypted data in section \ref{ER}.

\subsection{Genome-Wide Association and Linkage Studies}
In sections \ref{GB}-\ref{LD-section}, we provide more background about genomic data and LD test computation \cite{shahbazi2016private}.

\subsubsection{Genomic Background}
\label{GB}
\emph{DNA} is a sequence of nucleotides $\{A, C, G, T\}$. An individual's collection of genes is called a \emph{genotype} and the physically observable characteristics of an individual are called a \emph{phenotype}. A \emph{genetic marker} is defined as a gene or a DNA segment with a known locus (location) on a chromosome, which is typically used to help link an inherited disease with the responsible gene. Then a set of closely linked genetic markers found in one chromosome that tend to be inherited together is called a \emph{haplotype}. 

A \emph{single nucleotide polymorphism} (SNP) represents a common type of a genetic variation among people in a single nucleotide that occurs at a specific locus in a genome. One of a number of alternative forms of a gene at a given locus is called an \emph{allele}. The most common and least common alleles that occur in a given population are called \emph{major} and \emph{minor} alleles, respectively. We denote a major allele by a capital letter, e.g., $A$, and a minor allele by the corresponding lowercase letter, e.g., $a$. An individual inherits two alleles for each gene, one from each parent. If the two alleles are the same, the individual is \emph{homozygous} for that gene and is \emph{heterozygous} otherwise. Based on that information, we distinguish between the following categories: homozygous reference genotype, denoted as $AA$; heterozygous genotype, denoted as $Aa$; and homozygous variant genotype denoted as $aa$. We refer to the two alleles inherited for a particular gene as a genotype.

Let $N$ denote the total number of collected alleles in a pool of genes. We then use $N_A$ and $N_a$ to denote the number of major and minor alleles in the observed population, respectively. Similarly, $N_{AA}$, $N_{Aa}$, and $N_{aa}$ denote the number of gene variants of the type $AA$, $Aa$, and $aa$, respectively. They are used to compute values $N_A$ and $N_a$ as $$N_A = 2 N_{AA} + N_{Aa}, \qquad \qquad N_a = 2 N_{aa} + N_{Aa}.$$

In addition, an allele frequency is defined as the number of this allele in a certain locus in the observed population. In other words, we define major and minor allele frequencies $p_A$ and $p_a$ as $p_A = N_A/N$ and $p_a = N_a/N$, respectively. A genotype frequency can be defined analogously.

\subsubsection{Linkage Disequilibrium}
\label{LD-section}
Linkage disequilibrium is an important notion in population genetics that occurs when genotypes at two different loci are not independent of each other. In other words, LD is the non-random association of pairs of
alleles that often descend from a single ancestral chromosome. Consider two loci $A$ and $B$ with two alleles each ($A$, $a$, $B$, and $b$). There are 9 possible genotypes $AABB$, $AABb$, $AAbb$, $AaBB$, $AaBb$, $Aabb$, $aaBB$, $aaBb$, $aabb$, and there are four haplotypes $AB$, $Ab$, $aB$, $ab$. Let us use $N_{AB}, N_{Ab}, N_{aB},$ and $N_{ab}$ as the number of instances of each of the four haplotypes in the observed population.  Then, their population frequencies are computed as:
\begin{align*}
p_{AB} = \frac{N_{AB}}{N}, && p_{Ab} = \frac{N_{Ab}}{N}, &&
p_{aB} = \frac{N_{aB}}{N}, && p_{ab} = \frac{N_{ab}}{N}.
\end{align*}

When the alleles' frequencies are independent (i.e., we have linkage equilibrium), we expect that:
\begin{align*}
p_{AB} = p_A p_B, && p_{Ab} = p_A p_b, && p_{aB} = p_a p_B, && p_{ab} = p_a p_b.
\end{align*}
where, as before, $p_A = N_A/N$, $p_a = N_a/N$ and similarly $p_B = N_B/N$,
$p_b = N_b/N$, but now $N_A = N_{AB} + N_{Ab}$, $N_a = N_{aB} + N_{ab}$, $N_B
= N_{AB} + N_{aB}$, $N_b = N_{Ab} + N_{ab}$. However, if the alleles are in
LD, the formulas become:
\begin{align*}
p_{AB} = p_A p_B + D_{AB}, && p_{Ab} = p_A p_b - D_{AB}, &&
p_{aB} = p_a p_B - D_{AB}, && p_{ab} = p_a p_b + D_{AB}.
\end{align*}
The parameter $D_{AB}$ is called the coefficient of LD and can be computed as $D_{AB} = p_{AB} - p_A p_B$.

Chi-square statistics for the hypothesis $H_0$ of no disequilibrium (i.e., $D_{AB}=0$) is computed as:
\begin{equation}\label{eq:LD1}\nonumber
\chi^2_{A,B} = \frac{2N \cdot D^2}{p_A \cdot p_a \cdot p_B \cdot p_b} 
= \frac{2N \cdot (N \cdot N_{AB} - N_A \cdot N_B)^2}{N_A \cdot N_a \cdot N_B \cdot N_b}
\end{equation}
$H_0$ is rejected (i.e., LD is present) if $\chi_{A,B} ^2 $ exceeds a particular threshold or

\begin{equation}\label{eq:LD2}\nonumber
2N \cdot (N \cdot N_{AB} - N_A \cdot N_B)^2 > \chi^2_{A,B} \cdot N_A \cdot
N_a \cdot N_B \cdot N_b.
\end{equation}

\subsection{Cryptographic Tools}
We design our secure computation protocols using Homomorphic Encryption (HE) and Garbled Circuit (GC). Note that we can use any HE including additive HE and fully HE, but in here we are more interested in exploring fully HE (e.g., Paillier encryption as the additive HE and Lattice-based cryptography as fully HE). In the following we describe HE and GC briefly and then focus on details of the proposed solution.

\subsubsection{Homomorphic Encryption}
HE is a type of encryption that allows computation to be performed on encrypted data without revealing any information about the original data. In here, we use a specific type of HE where its key is defined in a public-key cryptosystem. This scheme is defined by three algorithms ($\sf Gen$, $\sf Enc$, $\sf Dec$), where $\sf Gen$ is a key generation algorithm that on input of a security parameter $1^\kappa$ produces a public-private key pair $(pk, sk)$; $\sf Enc$ is an encryption algorithm that on input of a public key $pk$ and message $m$ produces ciphertext $c$; and $\sf Dec$ is a decryption algorithm that on input of a private key $sk$ and ciphertext $c$ produces decrypted message $m$ or special character $\perp$ that indicates failure. For conciseness, we use notation ${\sf Enc}_{pk}(m)$ or ${\sf Enc}(m)$ and ${\sf Dec}_{sk}(c)$ or ${\sf Dec}(c)$ in place of ${\sf Enc}(pk, m)$ and ${\sf Dec}(sk, c)$, respectively. A semantically secure encryption scheme guarantees that no information about the encrypted message can be learned from its ciphertext with more than a negligible (in $\kappa$) probability.

Note that, in secure computation based on HE, the complexity of a protocol is measured based on non-free (expensive) operations. As an example, in additive HE, addition is a free operation and multiplication is counted as an expensive operation. Therefore, to optimize a solution we need to minimize non-free operations. We can also provide the complexity of a designed protocol based HE in terms of communication and computation complexities of no-free operations. While, fully HE supports arbitrary computation and it is a more powerful tool, but we need to define a specific noise budget for sequential multiplication operations which affects the performance of a computation. Since, we use the SEAL library for implementation, more information about fully HE can be found in \cite{sealcrypto}.   

\subsubsection{Garbled Circuit}
The use of GC allows two parties $P_1$ and $P_2$ to securely evaluate a Boolean circuit of their choice. That is, given an arbitrary function $f(x_1, x_2)$ that depends on private inputs $x_1$ and $x_2$ of $P_1$ and $P_2$, respectively, the parties first represent is as a Boolean circuit. One party, say $P_1$, acts as a circuit generator and creates a garbled representation of the circuit by associating both values of each binary wire with random labels. The other party, say $P_2$, acts as a circuit evaluator and evaluates the circuit in its garbled representation without knowing the meaning of the labels that it handles during the evaluation. The output labels can be mapped to their meaning and revealed to either or both parties.

The fastest currently available approach for circuit generation and evaluation we are aware of is by Bellare et al. \cite{bel13}. It is compatible with earlier optimizations, most notably the ``free XOR'' gate technique \cite{kol08} that allows XOR gates to be processed without cryptographic operations or communication, resulting in virtually no overhead for such gates. A recent half-gates optimization \cite{zah15} can also be applied to this construction to reduce communication associated with garbled gates. In addition, there are some recent works on GC compilers (e.g., \cite{tinygarble,compgc}) which are designed based on \cite{bel13}.

An important component of garbled circuit evaluation is 1-out-of-2 Oblivious Transfer (OT). It allows the circuit evaluator to obtain wire labels corresponding to its inputs. In particular, in OT the sender (i.e., circuit generator in our case) possesses two strings $s_0$ and $s_1$ and the receiver (circuit evaluator) has a bit $\sigma$. OT allows the receiver to obtain string $s_\sigma$ and the sender learns nothing. 

Note that, in the two-party setting solution based on GC, the complexity of an operation is measured in the number of non-free (i.e., non-XOR) Boolean gates because of optimization in XOR gate. Also, some computations like shift operation do not consist of any kind of gate and it is totally free. Therefore, to have an optimized solution, we need to minimize the number of non-XOR gates by using more free operations during the computation instead. In addition, we can report the complexity of a designed protocol in terms of the number of non-free gates.

\section{Designed Protocols}

In both of the following protocols, we assume we have access to Crypto Service Provider (CSP), who is a trusted third party with access to implementations of cryptographic standards and algorithms. (Such a CSP could possibly be added as a participant in future versions of the Computable protocol \cite{ComputableWhitepaper}) We also assume the presence of a datatrust (DT), makers $o_i$ where $i = 1, \ldots, n$, and buyers $s_j$ where $j = 1, \ldots, m$. At the end of protocol execution, each $s_j$ learns the result of a secure computation. 

\subsection{Homomorphic Encryption Protocol}
\begin{figure}[h!]
  \caption{Designed protocol based on HE with steps corresponding to those in Protocol 1.}
  \includegraphics[scale=0.7]{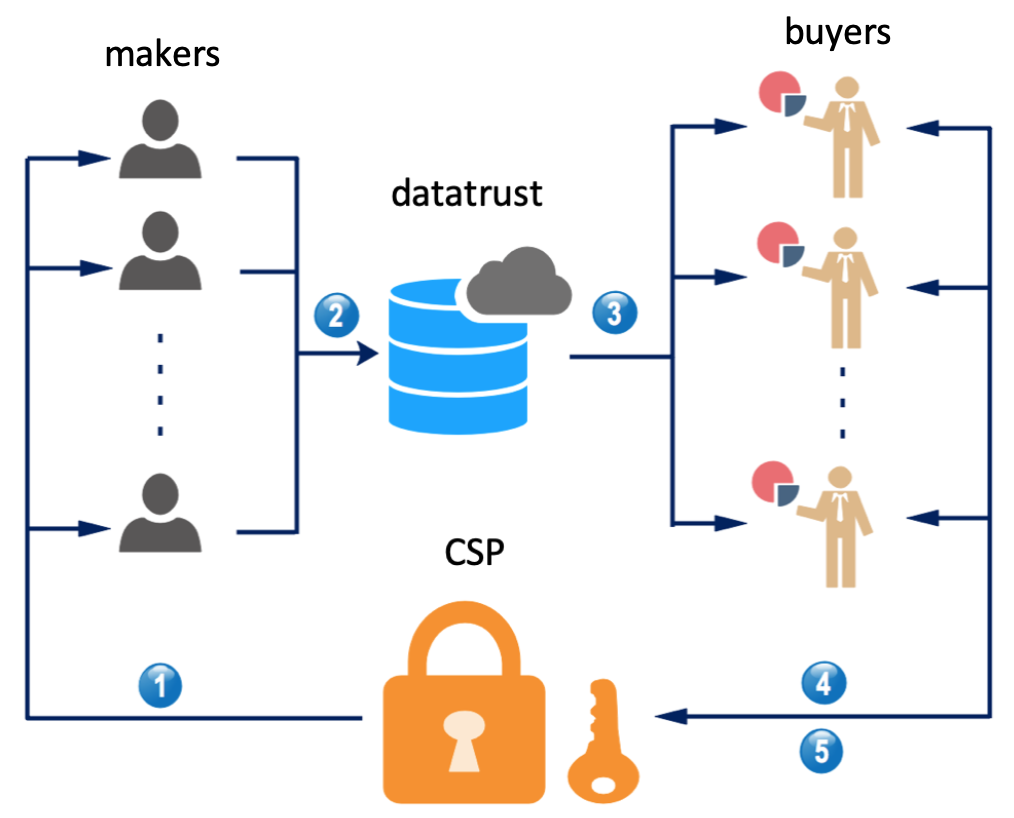}
  \label{hefig}
\end{figure}

In HE, we have access to its three main algorithms ($\sf Gen$, $\sf Enc$, $\sf Dec$). In this section, we use fully HE (FHE) developed by Brakerski/Fan-Vercauteren (BFV) and Cheon-Kim-Kim-Song (CKKS) as implemented by the SEAL library \cite{sealcrypto}. We introduce our solution in Protocol 1 and associated Figure \ref{hefig}.

\begin{figure}[h]
\fbox{\parbox{.97\columnwidth}{\small
\label{protocol1}
\textbf{Protocol 1:}\\
\textbf{Inputs:} Security parameter $\kappa$, a set of data $(x_1, \ldots, x_n)$ where $x_i$ belongs to maker $o_i$, and function $f$ which can be arbitrary computation (e.g., LR or LD tests).\\
\textbf{Outputs:} Each $s_j$ learns $f(x_1, \ldots, x_n)$.\\
\begin{enumerate}
\item CSP generates a public-private key pair $(pk, sk) \leftarrow {\sf Gen}(1^\kappa)$, and makes $pk$ available for everyone.

\item Each $o_i$ encrypts it own listing by computing $c_i = {\sf Enc}_{pk}(x_i)$, and submits $c_i$ to DT.

\item Each $s_j$ can send a query to DT to receive $(c_1, \ldots, c_n)$.

\item Each $s_j$ computes $C = f'(c_1, \ldots, c_n) = {\sf Enc}_{pk}(f(x_1, \ldots, x_n))$. $f'$ can be defined and performs by using homomorphic properties of the underlying HE scheme.

\item Each $s_j$ can send $C$ to CSP, and CSP computes $f(x_1, \ldots, x_n) = {\sf Dec}_{sk}(C)$ and sends the result to $s_j$.
\end{enumerate}}}
\end{figure}

\subsection{Garbled Circuit Protocol}
\begin{figure}[h!]
  \caption{Designed protocol based on GC with indicated steps.}
  \includegraphics[scale=0.7]{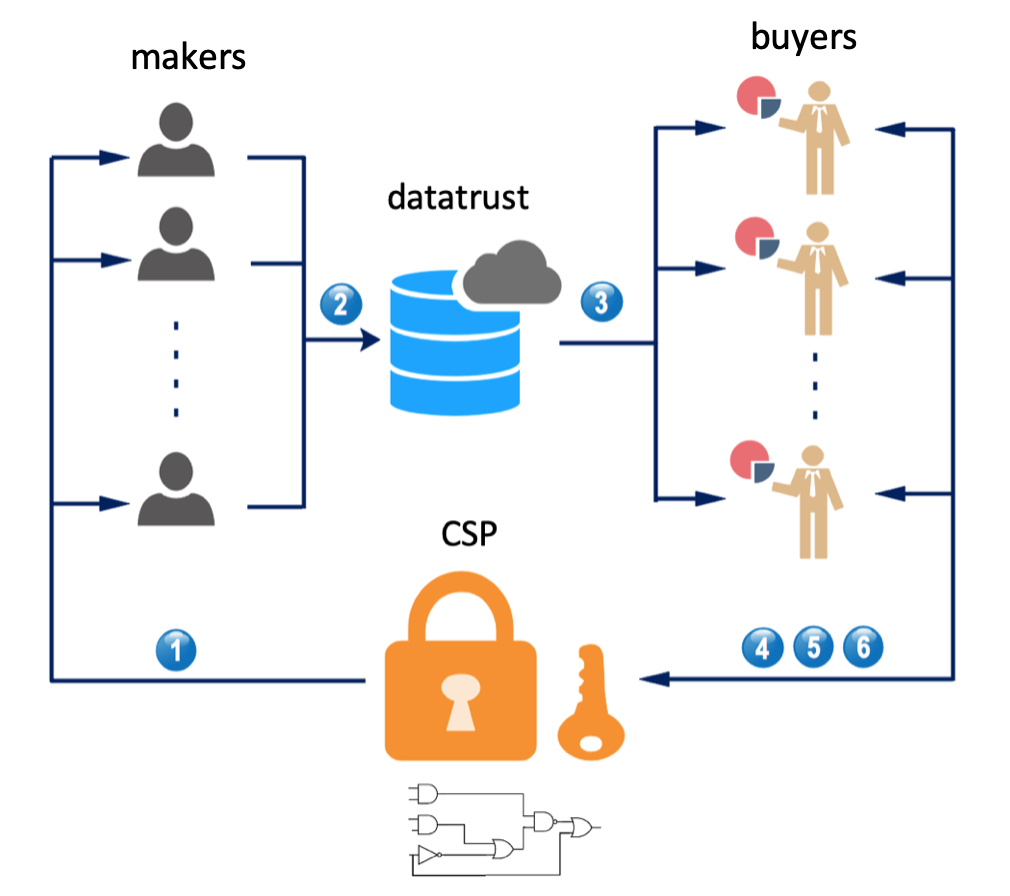}
  \label{gcfig}
\end{figure}

Next, we describe the details of the proposed solution based on GC. We have the same architecture as in Protocol 1, but instead of HE, GC is used as the underlying cryptographic tool. In this setting, the CSP needs to have enough computational power and storage to perform the garbling process. We introduce our solution in Protocol 2 and associated Figure \ref{gcfig}.

\begin{figure}[h]
\fbox{\parbox{.97\columnwidth}{\small
\textbf{Protocol 2:}\\
\textbf{Inputs:} A set of data $(x_1, \ldots, x_n)$ where $x_i$ belongs to maker $o_i$, a function $f$ which can be arbitrary computation (e.g., LR or LD tests), and a secure pseudorandom function $\sf PRF$.\\
\textbf{Outputs:} Each $s_j$ learns $f(x_1, \ldots, x_n)$.\\
\begin{enumerate}
\item CSP generates $\delta \stackrel{R}{\leftarrow} \{0, 1\}^{\kappa-1}$, $k \stackrel{R}{\leftarrow} \{0, 1\}^{\kappa}$, sets $\Delta = \delta||1$ (concatenation), and sends $\Delta$ and $k$ to $o_i$s.

\item Each $o_i$ computes wire labels $l_{(i, k)}^0 = {\sf PRF}(k, i||k)$ and $l_{(i, k)}^1 = l_{(i, k)}^0 \oplus \Delta$ for each bit $b_k$ of its own data, and sends all $l_{(i, k)}^{b_k}$s to DT.

\item Each $s_j$ can send a query to DT to receive all input wires.

\item CSP creates all input label pairs by using $\Delta$ and $k$ and creates circuit $C_f$ for any $f$. CSP sends circuit $C_f$ to $s_j$.

\item Each $s_j$ can evaluate circuit $C_f$ by using input labels, and sends output labels to CSP.

\item CSP sends the meaning of output labels to $s_j$.
\end{enumerate}}}
\end{figure}

\section{Experimental Results}
\label{ER}
In this section we evaluate the performance of our solution. The garbled circuit implementations were written in C and used the JustGarble library \cite{bel13,JG} for circuit garbling and evaluation. Our code supports the half-gates optimization \cite{zah15}. The FHE implementations were performed using the SEAL library \cite{sealcrypto}. All the computation for GC was run on a 3.3GHz machine, and for HE was run on a 2.7GHz machine, and experimental runs were repeated 10 times and mean values reported.

\subsection{Linkage Disequilibrium Results}
The GC protocol for the LD test results is reported in Table \ref{tab1}. Note that for the LD test, we vary the value of $N$ to demonstrate how this variable affects performance of the computation. Furthermore, we also vary the number $M$ of SNPs or alleles for which each test is run, with all $M$ instances of each test being executed at the same time.

\begin{table}[h]
\center
\begin{tabular}{ |c|c|c|c|c|c|c|} \hline
$M$ & $N$ & garbling & evaluation & \#gates & \#non-XOR gates & Comm.\\
\hline
\multirow{4}{*}{10} & 200 & 10.8  & 6.7 & 293550 & 81690 & 1.3\\ \cline{2-7}
& 400 & 14.1 & 7.7 & 313370 & 87150 & 1.4\\ \cline{2-7}
& 800 & 12.3 & 7.7 & 334510 & 92970 & 1.5\\ \cline{2-7}
& 1600 & 16.7 & 10.5 & 356970 & 99150 & 1.6\\ \hline \hline

\multirow{4}{*}{100} & 200 & 145.4 & 83.6 & 2935500 & 816900 & 13.1\\ \cline{2-7}
& 400 & 161.5 & 93.1 & 3133700 & 871500 & 14.0\\ \cline{2-7}
& 800 & 124.0 & 75.7 & 3345100 & 929700 & 14.9\\ \cline{2-7}
& 1600 & 132.9 & 80.2 & 3569700 & 991500 & 15.9\\ \hline \hline

\multirow{4}{*}{1000} & 200 & 1108.7 & 675.4 & 29355000 & 8169000 & 131.0\\ \cline{2-7}
& 4000 & 1269.6 & 758.7 & 31337000 & 8715000 & 139.7\\ \cline{2-7}
& 8000 & 1391.2 & 831.8 & 33451000 & 9297000 & 149.0\\ \cline{2-7}
& 16000 & 1371.4 & 897.6 & 35697000 & 9915000 & 158.9\\ \hline \hline
\end{tabular}
\caption{Execution time for LD test in ms and the communication in MB for GC.}
\label{tab1}
\end{table}

\begin{table}
\center
\begin{tabular}{ |c|c|c|c|} \hline
$M$ & Execution & Space & Expected execution (batch)\\
\hline
10 & 54.2 s & 18.34 MB & 1840 ms \\
\hline
100 & 9.1 m & 183.4 MB & 0.18 s \\
\hline
1000 & 1.48 h & 1.82 GB &  1.84 s \\
\hline
\end{tabular}
\caption{Execution time and space complexity for LD test for FHE.}
\label{tab2}
\end{table}

In addition, we implemented the LD test for the Brakerski/Fan-Vercauteren (BFV) scheme by using the SEAL Library \cite{sealcrypto}. Running the LD test takes $54.2$ seconds when $M = 10$, and runtime grows linearly with the size of $M$. Further details about the execution time are provided in Table \ref{tab2}. The SEAL library provides the facility to run HE operations in a batch. The LD test is very amenable to batch computation, and our execution becomes about 3000 times faster when all independent operations are run in a batch. Note that in our experiment, we set the polynomial modulus degree to 8192 and coefficient modulus to 128 and we reported the upper-bound of space complexity in Table \ref{tab2}. Note that in FHE, $M$ is the only LD test parameter that is important in the experiments because based on the selected parameters of FHE, the variable size of $N$ is covered. 

\subsection{Logistic Regression Results}
For the LR test, the computation becomes more complicated, since the exponentiation operation is not supported by the standard SEAL and JustGarble libraries. One potential solution to implement this operation is by using a private lookup table \cite{fingerprint}. In this approach we precompute the values of the exponential function for the desired precision and the range of input values and use private lookup to select the output based on private input. 

Consider an exponentiation function ($\sf Exp$) that needs to be evaluated on private input $a$ and in our case, it is defined over fixed-point arithmetic. Let the value of $a$ be in the range $[a_{min}, a_{max}]$ with $N$ denoting the number of the elements in the range. Then the approach consists of precomputing the function on all possible inputs and storing the result in an array $Z = (z_0, {\ldots}, z_{N-1})$. Consequently, evaluation of the function on private $a$ corresponds to privately retrieving the needed element of the array $Z$ using $a$ to determine the index. This procedure is formalized in the protocol $\sf Exp$ below. For further details, see reference \cite{fingerprint}. 

{\small \noindent \line(1,0){340} \\ [-0.02in]
$[b] \leftarrow {\sf Exp}([a], Z = \{z_i\}_{i=0}^{N-1})$ \\[-0.09in]
\line(1,0){340}
\begin{enumerate}
\item Compute $[b] \leftarrow {\sf Lookup}(\langle z_0, {\ldots}, z_{N-1}
  \rangle, [a])$.
\item Return $[b]$.
\end{enumerate}
\line(1,0){340}}

This approach can be implemented by using a multiplexer. However, this approach does not work well for FHE because its performance directly depends on the range of input values. For larger range, we need more sequential multiplications in the multiplexer, and as a result a larger noise budget ensues, making the solution less efficient. But the private lookup table is a reasonable solution for GC based protocols. In Table \ref{tab3}, we report performance of LR on GC (testing phase) on the breast cancer Wisconsin dataset \cite{dataset} where the size of inputs is 16 bits and we have different ranges for input values (in bits) for exponentiation operation. This dataset is a binary classification dataset with 30 dimensions and 569 sample data points. 
\begin{table}[h]
\center
\begin{tabular}{ |c|c|c|c|c|c|} \hline
range & garbling & evaluation & \#gates & \#non-XOR gates & Comm.\\
\hline
10 & 8.2 & 4.8 & 198909 & 106016 & 5.1\\ \hline
11 & 14.6 & 8.7 & 318717 & 193056 & 9.3\\ \hline
12 & 28.6 & 17.0 & 562429 & 371232 & 17.8\\ \hline \hline
\end{tabular}
\caption{Execution time for LR test in ms and the communication in MB for GC.}
\label{tab3}
\end{table}

\section{Conclusions and Future Directions}
In this paper, we design secure, efficient, and general protocols based on homomorphic encryption and garbled circuits to perform computation on sensitive encrypted data in a decentralized data market. We use examples from healthcare to emphasize the applicability of our protocol to sensitive datasets. The designed protocols are general and can be used for arbitrary computation, but we report performance only on our examples of linkage disequilibrium and logistic regression. To the best of our knowledge, our designed protocols are efficiently constructed. Our architecture is especially efficient for the garbled circuit protocol due to the fact that we eliminate oblivious transfer, the most computationally expensive part of GC. Our designed solutions are comparable and competitive with existing protocols including \cite{shahbazi2016private}.

In addition, our proposed solutions are theoretically salable for larger volumes of inputs, but achieving sufficient efficiency is challenging. More specifically, for lager inputs we may need to define more noise budget in HE protocol (operations that need to be done play an important role to define noise budget) to be able to do all computations with enough precision, and that may cause the solution inefficient in practice. Also, the performance of the private lookup table in GC protocol directly depends on the size of the table; therefore,  using the designed protocols for larger inputs in practice is not as straightforward as in theory. 

In the current work, the security of our design relies on the existence of an independent crypto service provider (CSP). The CSP is responsible for generating the public-private key pair in the HE scheme, and generates security parameters and garbles circuits in the GC scheme. In practice though, for many applications, we do not have access to such a trusted third party capable of acting as a CSP. As a future direction, we are working on a solution to eliminate the CSP and handle its role by performing a secure multi-party computation between the makers themselves. This approach may add some overhead to the protocols but it will make our design more broadly applicable for real-world use cases. 

Another major limitation of the current system is that each new computation requires a custom software implementation. For our experiments, we had to create custom code for both logistic regression and LD testing. Performing this implementation was nontrivial, and the computation of the exponent for logistic regression required some ingenuity. The construction of a more flexible software framework which can allow for broader classes of computation to be easily implemented is left to future work.

\bibliographystyle{abbrv}
\bibliography{main}

\end{document}